\documentclass[a4paper]{jpconf}
\usepackage{amssymb,latexsym,amscd,amsmath,graphicx,ifpdf}
\pdfoutput=1

\ifpdf

\DeclareGraphicsExtensions{.pdf}
\def\be{\begin{equation}}
\def\ee{\end{equation}}
\def\ba{\begin{array}}
\def\ea{\end{array}}
\def\bea{\begin{eqnarray}}
\def\eea{\end{eqnarray}}
\def\bi{\begin{itemize}}
\def\ei{\end{itemize}}

\def\half{{\textstyle{1\over2}}}

\begin{document}
\title{Constraints on the symmetry energy from neutron star observations}

\author{W. G. Newton$^1$, M. Gearheart$^1$, De-Hua Wen$^{1,2}$, and Bao-An~Li$^1$}

\address{$^1$ Department of Physics and Astronomy, Texas A\&M University - Commerce P.O. Box 3011, Commerce, TX, 75429-3011, USA}
\address{$^2$ Department of Physics, South China University of Technology,Guangzhou 510641, P.R. China}

\ead{William.Newton@tamuc.edu}

\begin{abstract}

The modeling of many neutron star observables incorporates the microphysics of both the stellar crust and core, which is tied intimately to the properties of the
nuclear matter equation of state (EoS). We explore the predictions of such models over the range of experimentally constrained nuclear matter 
parameters, focusing on the slope of the symmetry energy at nuclear saturation density $L$. We use a consistent model of the composition and EoS of neutron star 
crust and core matter to model the binding energy of pulsar B of the double pulsar system 
J0737-3039, the frequencies of torsional oscillations of the neutron star crust and the instability region for r-modes in the neutron star core damped by electron-electron viscosity at the crust-core interface. By confronting these
models with observations, we illustrate the potential of astrophysical observables to offer constraints on poorly known nuclear matter parameters complementary to terrestrial experiments, and demonstrate that our models consistently predict $L<70$ MeV.
\end{abstract}

\section{Introduction}

\begin{figure}[h]
\begin{center}
\includegraphics[width=36pc]{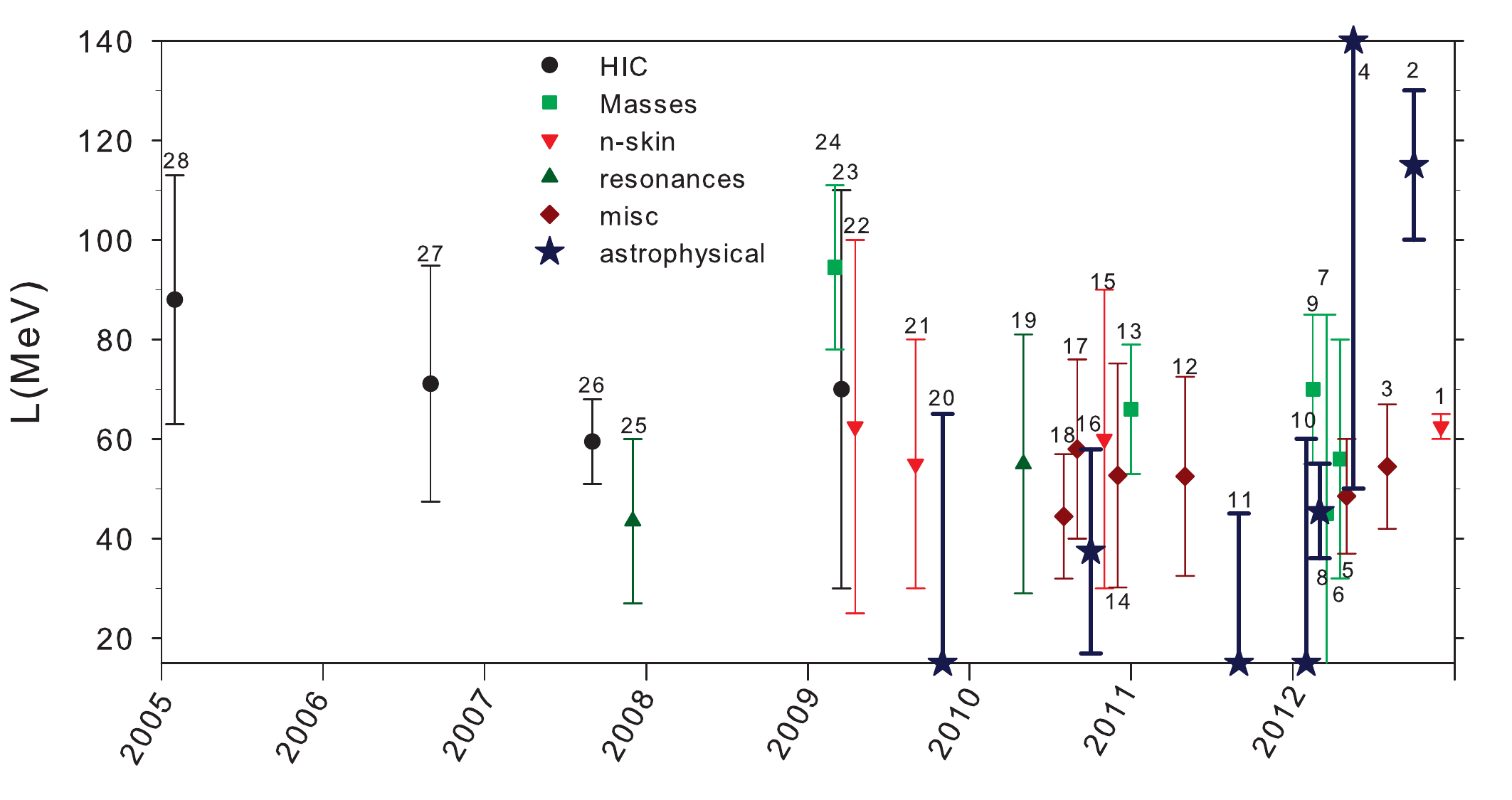}
\caption{\label{label1} Constraints on the slope of the symmetry energy at saturation density $L$ extracted from a variety of terrestrial experimental and astrophysical probes \cite{Agrawal:2012pq,Sotani:2012xd,Tews:2012fj,Vidana:2012ex,Piekarewicz:2012pp,Dong:2012zza,Lattimer:2012xj,Steiner:2011ft,Moller:2012aa,Wen:2011xz,Gearheart:2011qt,Chen:2011ek,Liu:2010ne,Xu:2010aa,Zenihiro:2010zz,Steiner:2010fz,Chen:2010qx,Hebeler:2010jx,Carbone:2010aa,Newton:2009vz,Warda:2009tc,Centelles:2008vu,Tsang:2008fd,Danielewicz:2008cm,Klimkiewicz:2007zz,Shetty:2007aa,Famiano:2006rb,Chen:2004si}. We show constraints obtained from modeling heavy ion collisions (HIC), nuclear masses, neutron (n-)skins of neutron rich nuclei, nuclear resonances, other nuclear experimental probes such as electrical polarizability and measurements of optical potentials (misc), and astrophysical observations including mass and radius measurements, torsional crust oscillations, the instability window for r-modes and the binding energy of PSR J0737-3039B.}
\end{center}\hspace{2pc}%
\end{figure}

The energy of an infinite system of nucleons $E(\delta, n_0)$ with proton fraction $x$ (or equivalently, isospin asymmetry $\delta = 1-2x$) and at a density around nuclear saturation density  $n_0 = 0.16$ baryons fm$^{-3}$ is a good approximation to the bulk energy inside terrestrial nuclei. Allowing density excursions $n$ further from saturation density (specified by the variable $\chi = \frac{n-n_{\rm 0}}{3n_{\rm 0}}$), $E(\delta, n)$ is the main ingredient in the calculation of the equation of state (EoS) and composition of neutron star matter. While this function is relatively well known in the regime most easily accessible to terrestrial experiment $\delta \ll 1, \chi \approx 0$, it is poorly constrained in density regimes far from saturation and for isospin asymmetries close to 1. These uncertain regimes of parameter space are where the study of the structure and collisions of neutron rich nuclei, and the description of astrophysical phenomena associated with neutron stars lie. Much effort has therefore been expended in constraining $E(\delta, n)$ (see Fig.~1).

A useful proxy for the uncertainties in the energy of isospin-asymmetric nuclear matter is the symmetry energy $S(n)$, defined by expanding $E(\delta, n)$ about $\delta = 0$ and $\chi = 0$:

\be\label{eq:eos1}
	E(\delta, n) = E_{\rm 0}(n) + S(n)\delta^2 + ...; \;\;\;\;\;\;\;\; S(n) = J + L \chi + \half K_{\rm sym} \chi^{2} + ...,
\ee
\noindent where $S(n)$ has the simple physical interpretation of being the difference between the energy of pure neutron matter (PNM) $E_{\rm PNM}(n)$ and symmetric nuclear matter (SNM) $E_0(n)$ when the expansion in $\delta$ is truncated to second order (the so-called parabolic approximation) $E_{\rm PNM}(n) \equiv E(n, \delta=1) \approx E_{\rm 0}(n) + S(n)$. The expansion in density around $n_0$ results in the definition of the symmetry energy at saturation density $J$, the symmetry energy slope parameter $L$ there, and the curvature $K_{\rm sym}$. Over the last few years, effort to constrain the asymmetric nuclear matter EoS has focussed on reducing the uncertainty in the slope parameter $L$, as is summarized in Fig.~1, which lists constraints on $L$ from a variety of terrestrial experimental probes extracted since 2005 (see also, e.g.,\cite{Tsang:2012se}). Since the properties of neutron stars are very much influenced by the behavior of the symmetry energy, various astrophysical observables sensitive to $L$ have been identified and used to extract constraints on $L$; these are also displayed in Fig.~1 by the thicker error bars denoted with stars. The mass and radius of a neutron star are the simplest properties to calculate, and confrontation with mass and radius measurements using recent data from X-ray bursts on accreting neutron stars have  resulted in some recent constraints on $L$ \cite{Steiner:2010fz,Steiner:2011ft}. In this proceeding, however, we shall focus on three different independent observables which, under certain interpretations, can be used to measure symmetry energy behavior: the gravitational binding energy of a neutron star, the frequency of torsional crust oscillations and the frequency at which core r-mode oscillations become unstable.

\begin{figure}[h]
\begin{center}
\includegraphics[width=32pc]{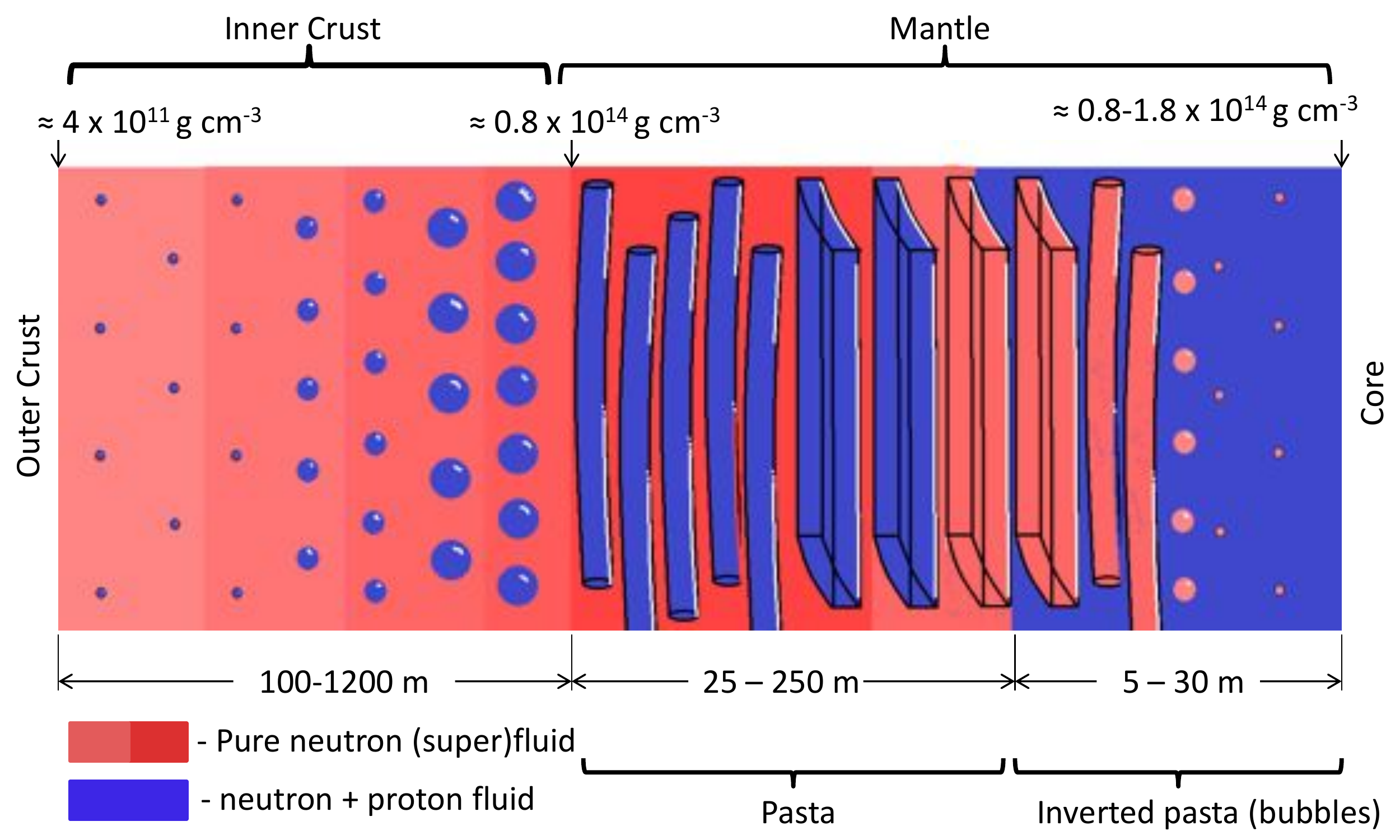}
\caption{\label{label} Illustration of the neutron star inner crust composition showing the lattice of nuclei surrounded by a neutron fluid and electrons in the lower density region, and the canonical sequence of exotic ``pasta'' nuclei appearing at in the higher density base of the crust: cylindrical, slab, inverted cylinder and spherical bubbles, the latter two referred to as the bubble phases. Included are ranges for the thickness of the whole inner crust, the whole of the pasta phases and the bubble region of the pasta phases, derived from our PNM sequence of crust and core EoSs over the slope of the symmetry energy range $25<L<115$ MeV and the neutron star mass range $1.0<M<2.0M_{\odot}$. Taken from \cite{Newton:2011aa}.}
\end{center}\hspace{2pc}
\end{figure}

The latter two observables involve interplay between the crust and core properties of the star: the overall size of the star as well as the thickness and mechanical properties of crustal matter affect the crust oscillation frequencies, while the r-mode oscillations have their primary viscous damping mechanism located at the crust-core interface for interior temperatures inferred from LMXBs. This is in contrast to observables such as mass and radius which have negligible dependence on uncertainties in the crust composition and EoS, and to extract constraints on the nuclear matter EoS from such observables requires modeling of crust and core matter properties using a consistent description of nuclear physics across both density regimes. We use a set of crust and core EoSs calculated using the same underlying Skyrme-like nuclear matter EoS $E(\delta, n)$ to calculate both crust and core composition, EoS and the transition densities between crust and core and between different phases of matter inside of the crust \cite{Newton:2011dw,Newton:2011aa}. We constrain $E(\delta, n)$ to reproduce the same underlying PNM EoS at low densities obtained from microscopic calculations \cite{Schwenk:2005ka,Gezerlis:2009iw,Gandolfi:2011xu,Hebeler:2009iv}, and vary $L$ smoothly over a conservative range $25 \lesssim L \lesssim 115$MeV while keeping the symmetric nuclear matter EoS fixed, thereby generating a set of EoSs we refer to as the PNM sequence. Ranges of predictions for the thickness of the inner crust and phases therein that emerge from our set of EoSs are shown in Fig.~2; note that we take into account not only the lattice of superheavy nuclei surrounded by a neutron fluid, but also the exotic nuclear shapes that might appear in the deepest layers of the crust, termed nuclear ``pasta''. Our set of crust and core EoSs and compositions are available for general use online \cite{Newton:2012online}.

\section{The gravitational binding energy of pulsar J0737-3039B}

The double pulsar system J0737-3039 \cite{Kramer:2008aa} has a low orbital eccentricity ($e=0.088$) \cite{Burgay:2003jj} and a low transverse velocity $v_{\rm t} \approx 10$ km/s \cite{Stairs:2006na}. Pulsar A has a relatively stable pulse profile and the angle between the spin axis of pulsar A and the orbital angular momentum of the system is small \cite{Breton:2008zz}. Together, these observations strongly suggest that the supernova that created the youngest of the two neutron stars, pulsar B, had little deviation from spherical symmetry. The most likely physical mechanism of a spherically symmetric supernova explosion involves the collapse of an ONeMg core of a progenitor star that started its life with a mass $\approx 8 - 10 M_{\odot}$ destabilized by electron ($e$)-captures onto Mg \cite{Podsiadlowski:2005ig}. Observationally, these supernovae would be classified as type Ib/c on account of the lack of observed Hydrogen lines in the spectra, the Hydrogen envelope of the progenitor having been removed prior to the supernova via binary interactions.  $e$-capture supernovae occur at a well defined core mass predicted to be $\approx 1.37 M_{\odot}$ \cite{Nomoto:1984aa,Podsiadlowski:2005ig,Kitaura:2005bt}. Additionally, simulations of this type of supernovae suggest very little mass loss from the core during the supernova, making $\approx 1.37 M_{\odot}$ an estimate of the \emph{baryon} mass of resultant neutron star, $M_{\rm B}$ \cite{Podsiadlowski:2005ig,Kitaura:2005bt,Newton:2009vz}.

\begin{figure}[t]
\begin{center}
\includegraphics[width=26pc]{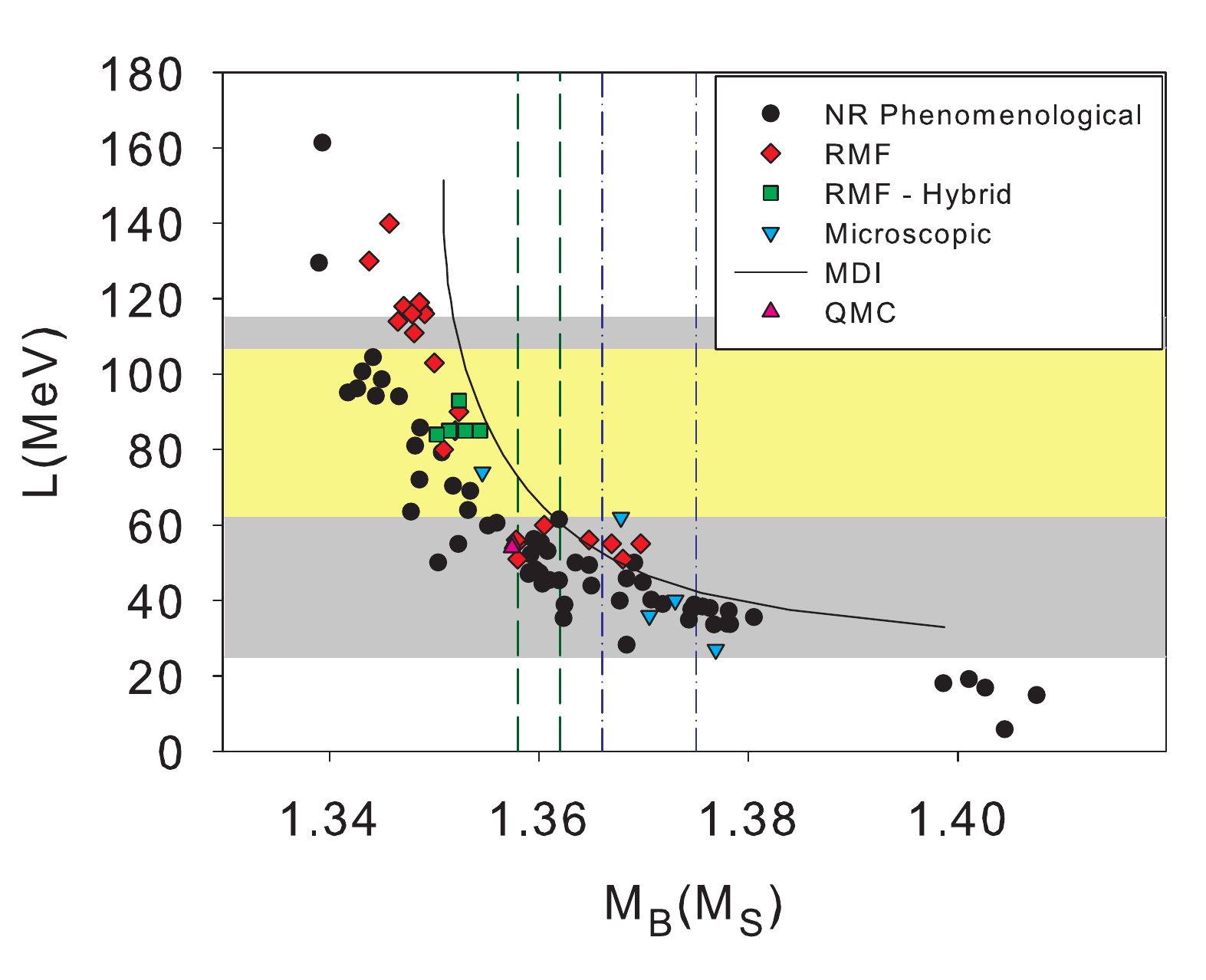}
\caption{\label{label}Baryon mass of an ONeMg core progenitor versus the slope of the symmetry energy at saturation density $L$ assuming that pulsar J0737-3039B (gravitational mass $M_{\rm G} = 1.2489 M_{\odot}$) was the result of a supernova induced by electron-captures onto Mg. The points are the predicted baryon masses from a wide range of EoSs including Skyrme and Gogny (MDI) non-relativistic (NR) phenomenological models, relativistic mean field (RMF) models with quark matter included (Hybrid EoSs) and without, microscopic nuclear models, and the quark-meson coupling model (QMC). The dark band is a conservative estimate of the aggregate constraint on $L$ from all nuclear experiments, while the lighter band shows the constraint on $L$ from isospin diffusion. The vertical dashed and dash-dotted lines represent predicted ranges for the baryon mass from two different studies of the progenitor evolution up to and through the supernova explosion taking into account estimates of mass loss during the explosion. Taken from \cite{Newton:2009vz}.}
\end{center}\hspace{2pc}%
\end{figure}

The mass of J0737-3039B has been accurately measured as $M_{\rm G} = 1.2489 \pm 0.0007M_{\odot}$ \cite{Kramer:2008aa}. This is the gravitational mass, the sum of the baryon mass and the gravitational binding energy in mass units $M_{\rm G} = M_{\rm B} + BE$. Neutron star EoSs predict a gravitational mass of $M_{\rm G} \sim 1.25 M_{\odot}$ for a star of baryon mass $M_{\rm B} \approx 1.37 M_{\odot}$, lending further credence to the $e$-capture supernova scenario as the formation mechanism of pulsar B \cite{Podsiadlowski:2005ig}. The exact gravitational binding energy, and hence $M_{\rm G}$, predicted by an EoS is a function of compactness $M_{\rm G}/R$ of the star and hence is sensitive to the symmetry energy slope at saturation density $L$. Under the assumption of the $e$-capture supernova scenario, we can therefore combine the progenitor modeling resulting in a prediction for $M_{\rm B}$ with the accurately measured $M_{\rm G}$ to obtain an astrophysical constraint on $L$ \cite{Newton:2009vz}.

Fig.~3 shows the baryon mass of a neutron star of gravitation mass $M_{\rm G} = 1.2489$ versus $L$ for a wide variety of EoSs including phenomenological models  (63 Skyrme parameterizations, 29 relativistic mean field parameterizations and the modified Gogny interaction MDI which allows for a smooth variation of $L$ while holding other nuclear matter parameters fixed), six EoSs based on microscopic calculations of the nucleon-nucleon interactions in medium, and two EoSs computed from the Quark-Meson coupling model (see \cite{Newton:2009vz} and references therein for details of these interactions). The vertical dashed lines give predicted ranges for the baryon mass from modeling the progenitor up to and through $e$-capture induced core-collapse, including effects such as mass loss from the core during explosion. The ranges taken together give the conservative astrophysical prediction of $1.358 < M_{\rm B} < 1.375M_{\odot}$, which are satisfied only by those EoSs which have $L<70$ MeV. By way of comparison, the light band shows the extracted range for $L$ from modeling isospin diffusion in heavy ion reactions   \cite{Chen:2004si} while the darker band shows a range of $L$ taking into account all experimental work shown in Fig.~1. For more details see \cite{Newton:2009vz}.

\section{Torsional crust oscillations}

Quasi-periodic oscillations observed in the X-ray light curves of three soft gamma-ray repeaters (SGRs) have found a popular interpretation as the torsional oscillation modes of the neutron star crust \cite{Israel:2005av,Strohmayer:2005ks,Strohmayer:2006py,Steiner:2009yg}. SGRs are believed to be highly magnetized neutron stars (magnetars, $B \sim 10^{15}$ G) which undergo occasional outbursts of soft gamma-ray and hard X-ray radiation. The most powerful flares are thought to be powered by magnetic reconnection analogously to solar flares, which requires large scale crust fracturing to allow the re-configuration of the magnetic field. The torsional crust oscillations are the post-fracturing ring-down phase.

\begin{figure}[h]
\begin{center}
\includegraphics[width=20pc]{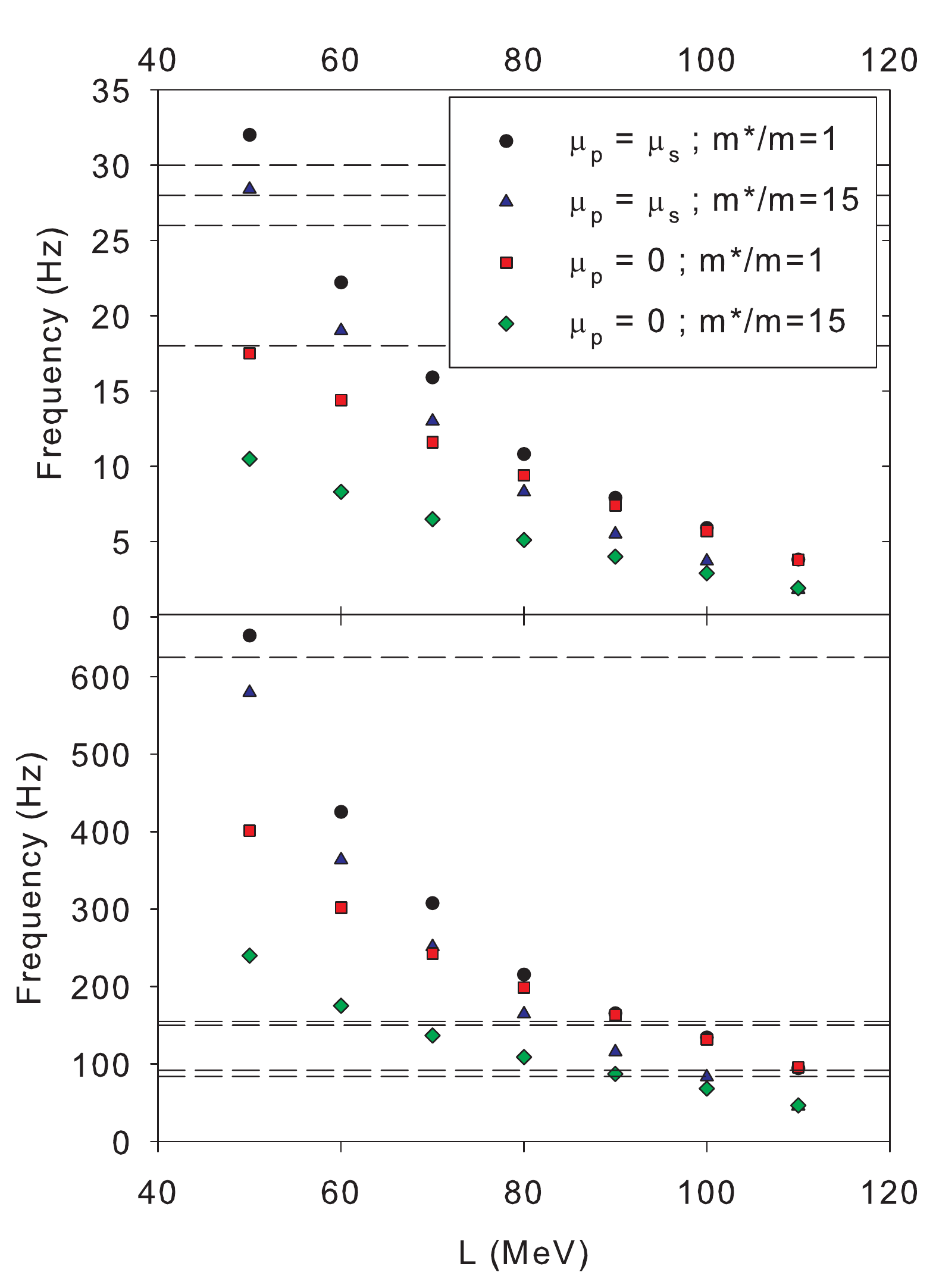}
\caption{\label{label}Fundamental (top) and first overtone (bottom) frequency of torsional crust oscillations versus the slope of the symmetry energy at saturation density $L$ for our set of EoSs with the shear modulus of pasta set to that of an elastic solid (circles and triangles) and set to zero (squares and diamonds), and taking the mesoscopic effective neutron mass to be equal to the bare mass (no entrainment; circles and squares) and equal to 15 times the bare mass (maximum entrainment - triangles and diamonds). The observed QPOs from SGR flares that fall in the displayed frequency ranges are shown by the horizontal dashed lines. Taken from \cite{Gearheart:2011qt}.}
\end{center}\hspace{2pc}%
\end{figure}

An approximate treatment of the frequencies of the fundamental and overtone modes of crust oscillations ($\omega_0$ and $\omega_n$ respectively) assuming an isotropic crust \cite{Samuelsson:2006tt} gives 

\be\label{freq}\begin{array}{l l}
	\displaystyle\omega^2_0 \approx \frac{e^{2\nu}v_s^2(l-1)(l+2)}{2RR_c},\;\;\;\; \omega^2_n \approx e^{\nu - \lambda} {n \pi v_s \over \Delta} \bigg[ 1 + e^{2\lambda} {(l-1)(l+2)\over 2 \pi^2} {\Delta^2 \over R R_c} {1 \over n^2},\bigg]
\end{array}\ee

\noindent where $n,l$ are the number of radial and angular nodes the mode has respectively. $M, R, R_{\rm c}$ and $\Delta$ are the stellar mass and radius, the radius out to the crust-core boundary and the thickness of the crust respectively. $v_{\rm s}$ is the shear speed at the base of the neutron star crust and $\nu$ and $\lambda$ are metric fields. $v_{\rm s}^2 = \mu/\rho$ where $\rho$ is the mass density and $\mu$ is the shear modulus at the base of the crust, given by \cite{Ogata:1990gm,Strohmayer:1991aa,Chugunov:2010ac}

\be \label{eq:shear_mod}
	\mu = 0.1106 \left(\frac{4\pi}{3}\right)^{1/3} A^{-4/3} n_{\rm b}^{4/3} (1-X_{\rm n})^{4/3} (Ze)^2.
\ee

\noindent Here, $A$ and $Z$ are the mass and charge numbers of the nuclei at the base of the crust (which occurs at a baryon density $n_{\rm b}$) and $X_{\rm n}$ is the density fraction of free neutrons there. The effects of the entrainment of the superfluid neutrons by the crustal lattice can be estimated by multiplying the frequencies by a factor of  $\epsilon_{\star} = { (1 - X_{\rm n}) /[1 - X_{\rm n} (m_{\rm n}^* / m_{\rm n})]}$ where $m_{\rm n}^*$ is the mesoscopic effective neutron mass \cite{Andersson:2008tp}. Here we will take $m_{\rm n}^* / m_{\rm n}$ = 1 (no entrainment) and $m_{\rm n}^* / m_{\rm n}$ = 15 (maximum entrainment) \cite{Chamel:2005aa}. To estimate upper and lower bounds on the effect of nuclear pasta on the frequencies, we  assuming (i) that the pasta has elastic properties similar to the rest of the solid crust, and therefore evaluate $\mu$ at the crust-core transition density $n_{\rm t}$, and (ii) that the pasta behaves as a fluid with no shear viscosity, effectively lowering the crust-core transition density to the density at which spherical nuclei make the transition to the pasta phases $n_{\rm p}$. Using our consistent set of EoSs of crust and core, we can calculate all of the above quantities with consistent nuclear matter parameters throughout crust and core over a range of values for $L$ \cite{Gearheart:2011qt}. 

The results for the fundamental and first overtone frequencies for a 1.4$M_{\odot}$ neutron star as a function of $L$ are plotted in Fig.~4 in the upper and lower panels respectively. The circles and squares give the frequencies neglecting the entrainment effect for the cases where the pasta behaves like the rest of the crust ($\mu_{\rm p} = \mu_{\rm s}$) and behaves like a fluid ($\mu_{\rm p}  = 0$) respectively; the upper limit on the effect of pasta is therefore to reduce the frequencies by a factor of $\approx 2$. The triangles and diamonds refer to the same two cases, but with the entrainment effect taken into account. This reduces the frequencies by about 20-30\% at most, and so is a somewhat smaller effect than that of the pasta phases. The horizontally dashed lines indicate the measured frequencies from SGRs. The requirement for our model to match any of the observed frequencies in the fundamental range constrains the slope of the symmetry energy to $L \lesssim 60$ MeV; if, additionally, say, the observed 625Hz mode is to be matched to the 1st overtone,  $L \lesssim 60$ MeV \emph{and} the pasta phases should have mechanical properties approaching that of an elastic solid. Of course, with a limited number of observations at our disposal, and many uncertainties remaining in the interpretation of the observations and the theoretical model of the oscillation modes, one must be wary of drawing any definitive conclusions from this work. For example, an analysis using a very similar model, but a different interpretation of which observed frequencies match which theoretically derived overtones, concludes that $100\lesssim L \lesssim 130$ MeV \cite{Sotani:2012xd}. Nevertheless, these analyses highlight a potentially promising astrophysical method for constraining the behavior of the symmetry energy.


\section{The r-mode instability and the maximum rotation rate of millisecond pulsars}

A growing number of pulsars ($\sim 150$ at present time) are observed to be rotating at periods of order milliseconds (ms). Such stars are presumed to be old, having undergone (or still undergoing) accretion-induced spin up via interaction with a binary companion. The maximum rotation frequency of any known pulsar is 716 Hz \cite{Hessels:2006ze}. However, a reasonable range of neutron star EoSs give theoretical maximum spin rates of $\sim 2000$ Hz and above (beyond which material will be ejected from the equator of the star). This raises the question: is there a physical mechanism limiting neutron star spin-up, or have we just got unlucky with our currently observed sample?

One possible physical mechanism concerns the stability of a class of inertial oscillation modes called r-modes. Such modes have as their restoring force the Coriolis force, and therefore only exist in rotating stars. They are closely related to Rossby waves in Earth's atmospheres and oceans. Above a certain rotation frequency, r-modes can become unstable through the Chandrasekhar-Friedman-Schutz (CFS) mechanism \cite{Chandrasekhar:1970aa,Friedmann:1978aa} in which gravitational radiation from the modes drives the oscillations to higher amplitudes. Such radiation dissipates rotational kinetic energy and the star spins down on short timescales. Thus if this instability occurs at the right frequency, it could explain why we see no stars rotating with greater frequencies \cite{Bildsten:1998ey,Andersson:1998qs}.

The onset of the instability is determined by the competition between the core viscosity which damps the mode, and the driving force caused by the gravitational radiation. By equating the  gravitational radiation and viscous damping timescales, given by \cite{Lindblom:1998wf, Lindblom:2000gu, Owen:1998xg}

\begin{equation}\label{TGR}
 {1\over\tau_{GR}} =  {32\pi G \Omega^{2l+2}\over c^{2l+3}}
{(l-1)^{2l}\over [(2l+1)!!]^2}  \left({l+2\over
l+1}\right)^{2l+2} \int_0^{R_c}\rho r^{2l+2} dr,
\end{equation}

\begin{equation}\label{TV}
 \tau_v  = \frac{1}{2\Omega} \frac{{2^{l+3/2}(l+1)!}}{l(2l+1)!!{\cal
 I}_l}
\sqrt{2\Omega
R_c^2\rho_c\over\eta_c} \int_0^{R_c}
{\rho\over\rho_c}\left({r\over R_c}\right)^{2l+2} {dr\over R_c},
\end{equation}

\noindent respectively, we can solve for the frequency of onset of the CFS instability. Here, the viscosities are evaluated at the crust-core boundary \cite{Levin:2001aa} where the r-mode amplitude is largest. $\rho_{\rm c}$ is the crust-core transition density and $R_{\rm c}$ the stellar radius at that density and $\nu_{\rm c}$ the viscosity there. We consider the case of $l=2$, with $\mathcal{I}_2$ = 0.80411. For the temperatures estimated in several ms pulsars, the dominant source of viscosity is given by the electron-electron scattering process $ \eta_{ee}=6.0\times 10^6 \rho^2 T^{-2}$~(g cm$^{-1}$ s$^{-1})$ \cite{Flowers:1979aa,Cutler:1987aa}. By employing our consistent set of crust and core EoSs, we can calculate the frequency of onset of the CFS instability as a function of temperature for different values of $L$: the results of such calculations are plotted in Fig.~5 for a 1.4$M_{\odot}$ and 2.0$M_{\odot}$ star (left and right panels respectively) \cite{Wen:2011xz}. Since ms pulsars are known to have been accreting material for a large portion of their lives, we will concentrate on the 2.0$M_{\odot}$ results.

\begin{figure}[t]
\begin{center}
\includegraphics[width=18pc]{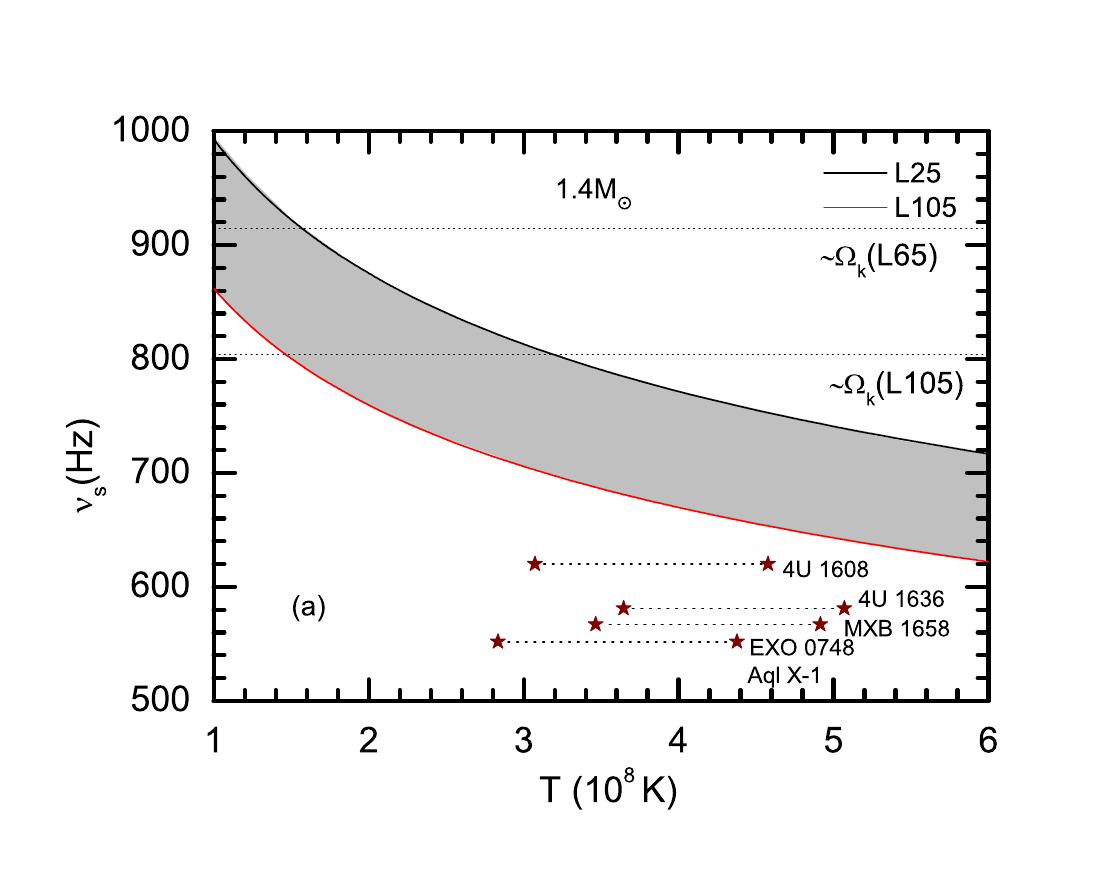}
\includegraphics[width=18pc]{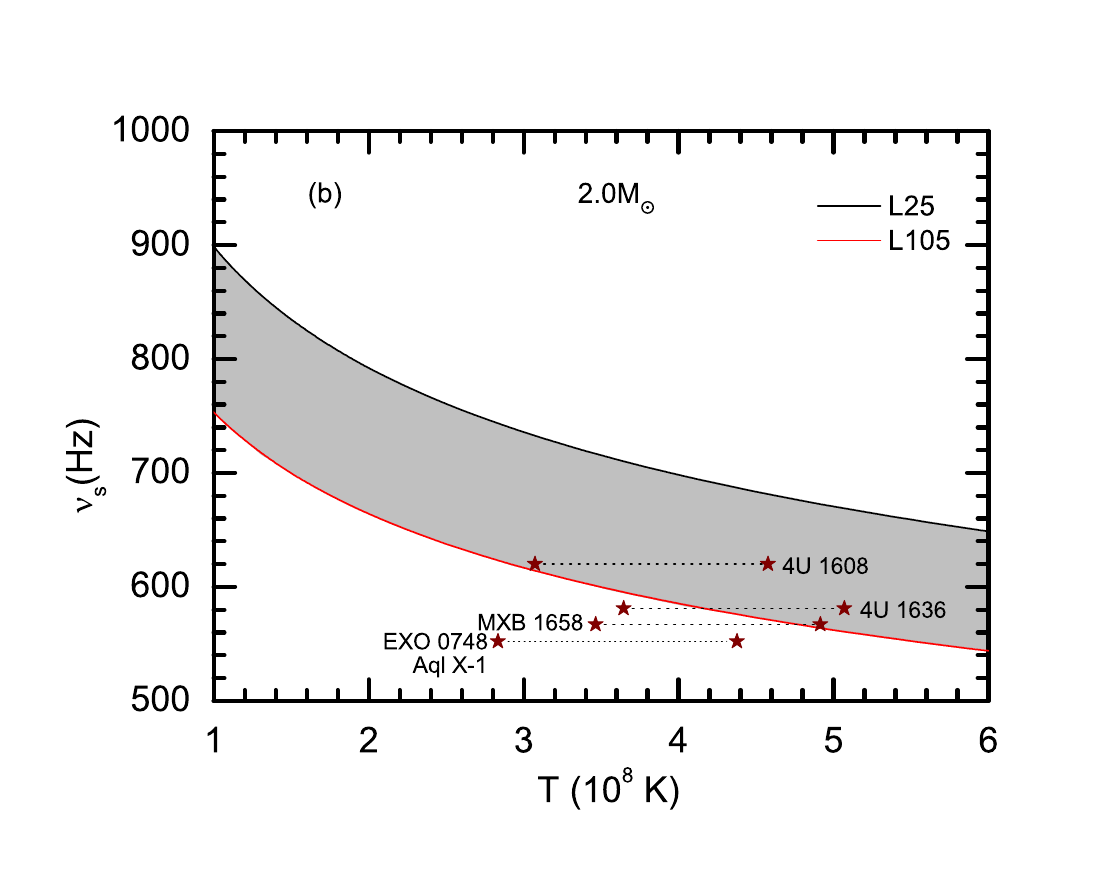}
\caption{\label{label}Frequency above which the electron-electron viscosity is insufficient to damp the gravitational radiation-driven r-mode instability versus core neutron star temperature for a 1.4$M_{\odot}$ star (left) and 2.0$M_{\odot}$ (right). Results are shown for the two bounding EoSs in our PNM sequence: $L=25$MeV and $L=115$MeV, with the region in between shaded in grey. The positions of 4 millisecond pulsars are shown with two estimates of their internal temperature; one interpretation of observations is that the frequency of onset of the r-mode instability should be higher than any observed pulsar frequency at a given temperature in order to be consistent with those observations. For comparison, the neutron star break-up frequency for $L=25, 65$ and $105$ MeV are shown if they fall within the frequency range of the plot. Taken from \cite{Wen:2011xz}.}
\end{center}
\end{figure}

The solid curve shows the onset frequency for $L$=25 MeV; the dotted line for $L$=115 MeV. The location of four observed ms pulsars in short recurrence-time LMXBs are shown by the star symbols; the temperature has been estimated in two different ways \cite{Watts:2008qw,Keek:2010xx}. As an example of how this analysis might constrain the symmetry energy, consider the object 4U 1608. For either estimate of temperature, this object is above the instability line for $L=105$ MeV; hence, given all other model assumptions, this constrains $L$ to be somewhat lower than 105 MeV. Indeed, under the assumption that the r-mode instability is the limiting mechanism for neutron star spins, and that our model is correct, one is forced to conclude that $L<65$ MeV in order for the higher temperature estimate of 4U 1608 to be consistent with our results.

Again, there is still much uncertainty regarding the limiting spin mechanism, and whether it is even necessary to invoke an extra physical mechanism to explain observations \cite{Ho:2011tt}. There is additional uncertainty in the theoretical modeling of r-modes. Indeed, a similar analysis to ours concludes that $L>50$ MeV \cite{Vidana:2012ex}. This analysis, however, differs from ours in that it takes into account viscous dissipation throughout the whole core instead of only at the crust-core boundary layer, and it does not consistently model the neutron star core and crust together: the radius is varied independently of both, even though the radius strongly correlated with crust thickness and core composition. Nevertheless, both studies demonstrate the sensitivity of the r-mode instability window to $L$ and offer a potential additional astrophysical observable to constrain it.

\section{Conclusions}

Most discussion of obtaining astrophysical constraints on the density-dependent behavior of the symmetry energy of nuclear matter centers around measuring simple global neutron star properties such as mass and radius for which the theoretical modeling is relatively simple even though observational limits on radius, in particular, remain somewhat uncertain. We have reviewed three different, and mutually independent, observations relating to neutron star systems and explored their ability to provide constraints on the slope of the symmetry energy at saturation density $L$ complementary to those obtained from terrestrial nuclear experiments and theoretical calculations. 

Unlike the observations of radius, for example, the observed mass of J0737-3039B, the frequencies of QPOs from 3 SGRs and the spin frequencies of ms pulsars are all observed to a high degree of accuracy. The uncertainties instead lie in the interpretation of the observations linking, for example, the mass of J0737-3039B to the binding energy of the star through an assumed evolutionary scenario for which only circumstantial evidence exists, or the observed QPOs to the torsional crust oscillations. On top of these uncertainties, the theoretical modeling of the underlying physical phenomena such as oscillation modes in neutron stars is much more challenging than calculating static global properties. Finally, on the nuclear physics side, the phenomena discussed here often involve interplay between the neutron star crust and core requiring knowledge not only of the EoS but of the crust and core composition, the size of each component, mechanical and transport properties and more. In order to extract information about the underlying nuclear EoS, all such microphysics must be modeled using a consistent nuclear matter EoS at all densities. 

In this proceeding we have discussed our work on improving the consistency of the nuclear microphysics by constructing a set of consistent crust and core EoSs that span a conservative range in their values of $L$ while conforming to our best experimental and theoretical knowledge of the symmetric nuclear matter and pure neutron matter EoS. By applying such EoSs to simple models of the gravitational binding energy of J0737-3039B, the frequencies of torsional crustal oscillations and the frequency of onset of the CFS instability, we are able to confront the observations and extract constraints of $L<70$ MeV, $L<60$ MeV and $L<65$ MeV respectively. Although the theoretical interpretations of the observations remain uncertain, and the astrophysical modeling is almost certainly over-simplified, it is striking that all three constraints are consistent with each other despite their disparate origins. Other similar analyses conclude slightly differently to us due to differences in interpretation of observations and in the astrophysical modeling. However, we have shown that combining information from a variety of neutron star observables will allow ever more stringent constraints to be placed on the nuclear EoS, and offers additional consistency checks on astrophysical modeling.

\ack
This work is supported in part by the National Aeronautics and Space Administration under grant NNX11AC41G issued through the Science Mission Directorate, and the National Science Foundation under Grants No. PHY-1068022 and No. PHY-0757839. D.H. Wen is also supported in part by the National Natural Science Foundation of China under Grant No.10947023,  and the Fundamental Research Funds for the Central University, China under Grant No.2012zz0079. The project is sponsored by SRF for ROCS, SEM.

\section*{References}
\bibliographystyle{iopart-num}
\bibliography{bibliography}

\end{document}